\documentclass[vecphys,letter]{svmult}

\usepackage{graphicx}        
\usepackage{multicol}        
\usepackage[bottom]{footmisc}

\begin{document}

\title*{New Model Atmospheres: Testing the Solar Spectrum in the UV}
\titlerunning{New Model Atmospheres} 

\author{L.H. Rodriguez-Merino\inst{1},
O. Cardona\inst{1},
E. Bertone\inst{1}, 
M. Chavez\inst{1,2} 
\and A. Buzzoni\inst{3}}

\authorrunning{Rodriguez-Merino et al.} 
\institute{Instituto Nacional de Astrof\'isica, \'Optica y Electr\'onica, 
Luis Enrique Erro 1, 72840, Tonantzintla, Puebla, M\'exico 
(\texttt{lino@inaoep.mx}, \texttt{ebertone@inaoep.mx}, 
\texttt{ocardona@inaoep.mx}, 
\texttt{mchavez@inaoep.mx}),
\and IAM- Instituto de Astronom\'ia y Meteorolog\'ia, Universidad de Guadalajara, 
Vallarta 2602, 44130, Guadalajara, M\'exico, 
\and INAF-Osservatorio Astronomico di Bologna, Via Ranzani 1, 40127, Bologna, 
Italy \texttt{buzzoni@bo.astro.it}}

\maketitle

\begin{abstract}
We present preliminary results on the calculation of synthetic spectra 
obtained with the stellar model atmospheres developed by Cardona, Crivellari,
and Simonneau. These new models have been used as input within the {\sc Synthe} series 
of codes developed by Kurucz. As a first step we have tested if {\sc Synthe} is able 
to handle these models which go down to $\log{\tau_{\rm Ross}}= -13$. We have successfully 
calculated a synthetic solar spectrum in the wavelength region 2000--4500~\AA\ at high 
resolution ($R=522\,000$). Within this initial test we have found that layers at optical 
depths with $\log{\tau_{\rm Ross}} < -7$ significantly affect the mid-UV properties of 
a synthetic spectrum computed from a solar model. We anticipate that these new extended 
models will be a valuable tool for the analysis of UV stellar light arising from the 
outermost layers of the atmospheres.
\end {abstract}

\section{Introduction}
\label{sec:1}
A set of spectral energy distributions (SEDs) is a very useful tool to analyze stellar spectra 
and the integrated spectral properties of stellar systems (via some evolutionary population
synthesis code, see e.g. Buzzoni 1995). Observational atlases of stellar SEDs generally lack of 
homogeneous and complete coverage of the main stellar parameters ($T_{\rm eff}, \log{g}$ and [M/H]), 
therefore many of the recent population analyses rely on results of stellar atmosphere modelling, 
a practise that has been eased with the development of faster computers and sophisticated computational 
codes. In order to calculate a theoretical stellar SED it is necessary to have a model atmosphere 
which describes the physical quantities at different depths and, ideally, a complete set of opacities 
that account for the absorption of the radiation passing throughout the atmosphere.

During the last decades several groups  have developed computational codes capable to calculate model 
atmospheres and spectra at high resolution, some of whom have allowed the public use of their codes, 
among others, the codes {\sc Atlas9}, {\sc Atlas12} and {\sc Synthe} built by 
Kurucz\footnote{http://kurucz.harvard.edu/} (1993a,b) and {\sc Tlusty} and {\sc Synspec} constructed by
Hubeny \& Lanz\footnote{http://nova.astro.umd.edu/} (1992). In particular, {\sc Atlas9} and {\sc Synthe} 
codes have been used by our group to calculate the UVBLUE grid of theoretical SEDs and to investigate 
its potential for stellar and populations studies in the ultraviolet (UV) wavelength interval (see 
Rodriguez-Merino et al. 2005 for more details).

The UV wavelength range has historically been challenging in many branches of modern astrophysics 
since the observed stellar spectra are not well reproduced by predictions of theoretical models. 
One possible reason is the missing opacity problem (see Holweger 1970; Gustafsson et al. 1975), 
but another probable reason is actually that most of the model atmospheres do not provide the 
atmosphere structure near the stellar surface, where most of the UV radiation emerges. Therefore, 
it is crucial to calculate new models which describe the outermost layers of the stellar atmospheres. 

In this work we briefly describe the structure of a new model atmosphere for the Sun  which 
incorporates layers down to $\log{\tau_{\rm Ross}}=-13$. That is, optical depths more than five 
orders of magnitude thinner compared to classical models currently in use. This model has been 
couplet to {\sc Synthe} codes for testing their compatibility and exploring the effects of such 
layers on the UV flux.

\section{Model Atmospheres}
\label{sec:2}

\begin{figure}[t]
\centering
\includegraphics[width=0.8\hsize]{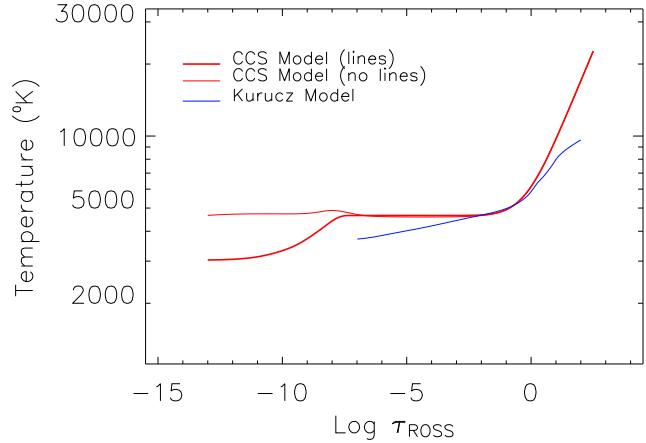}
\caption{The temperature profile of the CCS model atmosphere, with ({\it thick
curve}) and without ({\it thin curve}) line opacity, and of a Kurucz model ({\it dotted curve}) for 
solar physical parameters.}
\label{fig:1}
\end{figure}

The model atmospheres employed here have been developed by Cardona, Crivellari and Simonneau (CCS), 
these models use the Implicit Integral Method to solve the radiative transfer (Crivellari \& 
Simonneau 1994; Simonneau \& Crivellari 1993; Crivellari et al. 2002); this algorithm allows a 
stable and precise computation down to very low values of $\tau_{\rm Ross}$.  The models are based on approximations 
typical of classical models, namely, plane-parallel and homogeneous layers, steady state, hydrostatic, 
local thermodynamical, and radiative equilibria (convection has not been included here). We have 
considered the continuum opacity of ten elements (H, He, C, N, O, Na, Mg, Al, Si, Ca) and for three 
of them also the line opacity accounting for 20 absorption lines (12 for H, 6 for He, and 2 for Ca), 
which provide the large part of the absorption needed to compute the atmospheric structure. 
Figure~\ref{fig:1} displays a comparison between the temperature profile of the new model atmosphere 
with and without line opacity ({\it solid thick} and {\it thin lines}, respectively) and a Kurucz's 
model ({\it dotted line}) with similar physical parameters ($T_{\rm eff} = 5780$~K, $\log{g}= 4.5$, 
[M/H]=+0.0).

\begin{figure}[t]
\centering
\includegraphics[width=0.85\hsize]{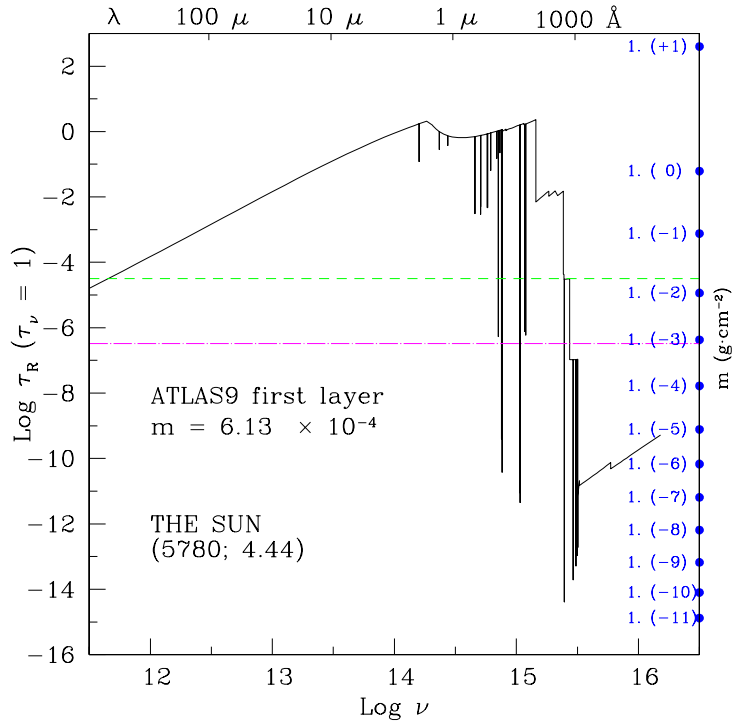}
\caption{Rosseland optical depth of formation of the main absorption features in a solar model as 
a function of frequency. The dot-dashed line displays the upper layer of the Kurucz model. Note that 
the UV features form well above the limiting log $\tau_{ROSS}\sim$ -7.}
\label{fig:2}
\end{figure}

In order to explore qualitatively the role of the most external atmospheric layers, in Fig.~\ref{fig:2} 
(kindly provided by L.~Crivellari) we track the value of the Rosseland optical depth at which the 
atmosphere becomes optically thin at each frequency. We can see how several strong lines and, in 
particular, the Lyman lines and break, which dominate the UV range, are produced in the most external 
layers ($\log{\tau_{\rm Ross}} \sim -7$ to $\log{\tau_{\rm Ross}} \sim -13$). Currently, there are no 
classical model atmospheres that take into account the effects of these external regions in the SEDs 
of stellar models.

\section{Synthetic Spectra}
\label{sec:3}
\begin{figure}[t]
\centering
\includegraphics[width=1\hsize]{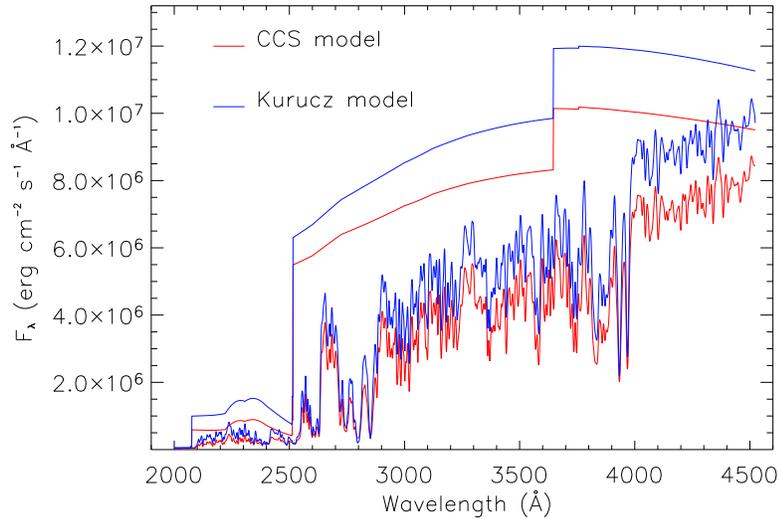}
\caption{Near-UV synthetic solar spectra computed with {\sc Synthe} using the CCS model with
line opacity and Kurucz model atmospheres. Spectra are shown at a resolution of FWHM=6~\AA.}
\label{fig:3}
\end{figure}

Once a model atmosphere has been computed the next step is to calculate the synthetic 
spectrum by solving the transfer equation at every layer in the atmosphere. Since the 
model here presented was not calculated within the Kurucz machinery, our main goal was 
to test if {\sc Synthe} is able to handle extended models, and if so, run {\sc Synthe} 
with the new model and analyze the effects of the layers at depths in the interval 
$\log{\tau_{\rm Ross}} \sim -7$ -- $\log{\tau_{\rm Ross}} \sim -13$ on the emergent 
spectrum. After successfully modifying the output of the solar CCS model with line opacity 
to be compatible with the input data required by {\sc Synthe}, we computed a synthetic 
spectrum at high resolution ($R=\lambda/\Delta\lambda=522\,000$) covering the near ultraviolet 
interval, from 2000 to 4500 \AA. The line list by Kurucz (1992) has been adopted to 
account for individual line absorption. In Figure~\ref{fig:3} we display the spectra 
(pure continuum and continuum + line absorption) obtained with CCS model ({\it solid line}) 
and a Kurucz's solar model ({\it dotted line}). For the sake of clarity, line spectra have 
been degraded with a Gaussian kernel to a resolution of FWHM=6~\AA. Although we are well aware 
that at present it is not possible to carry out any detailed comparison between the
calculated SED of the new model with either results from other codes or observed data 
(mainly due to the lack of line opacity and convective transport), it is interesting to note 
that the CCS model fluxes are sistematically lower in the wavelength interval employed.

\begin{figure}[t]
\centering
\includegraphics[width=0.49\hsize,height=7.5cm]{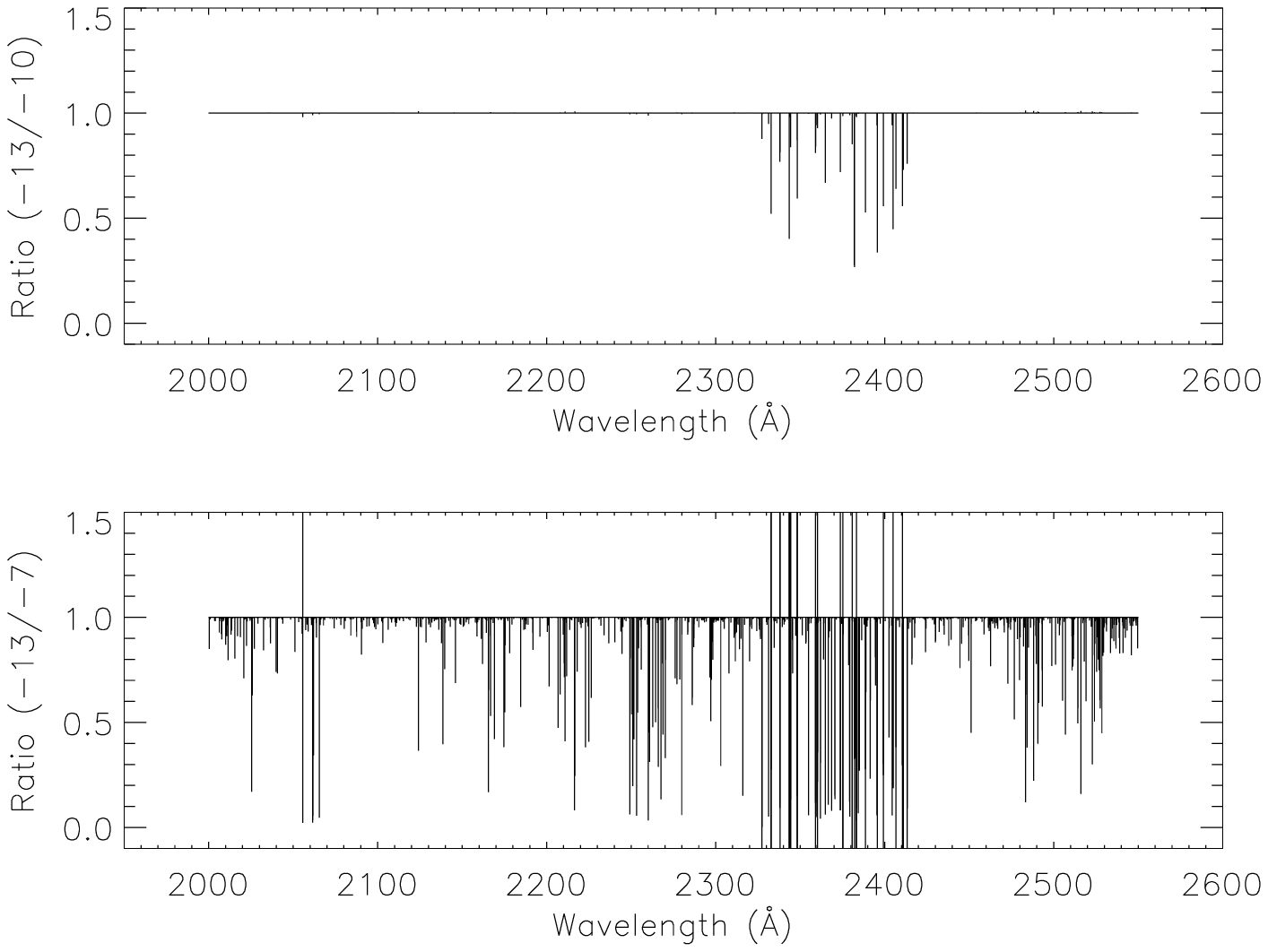}
\includegraphics[width=0.49\hsize,height=7.5cm]{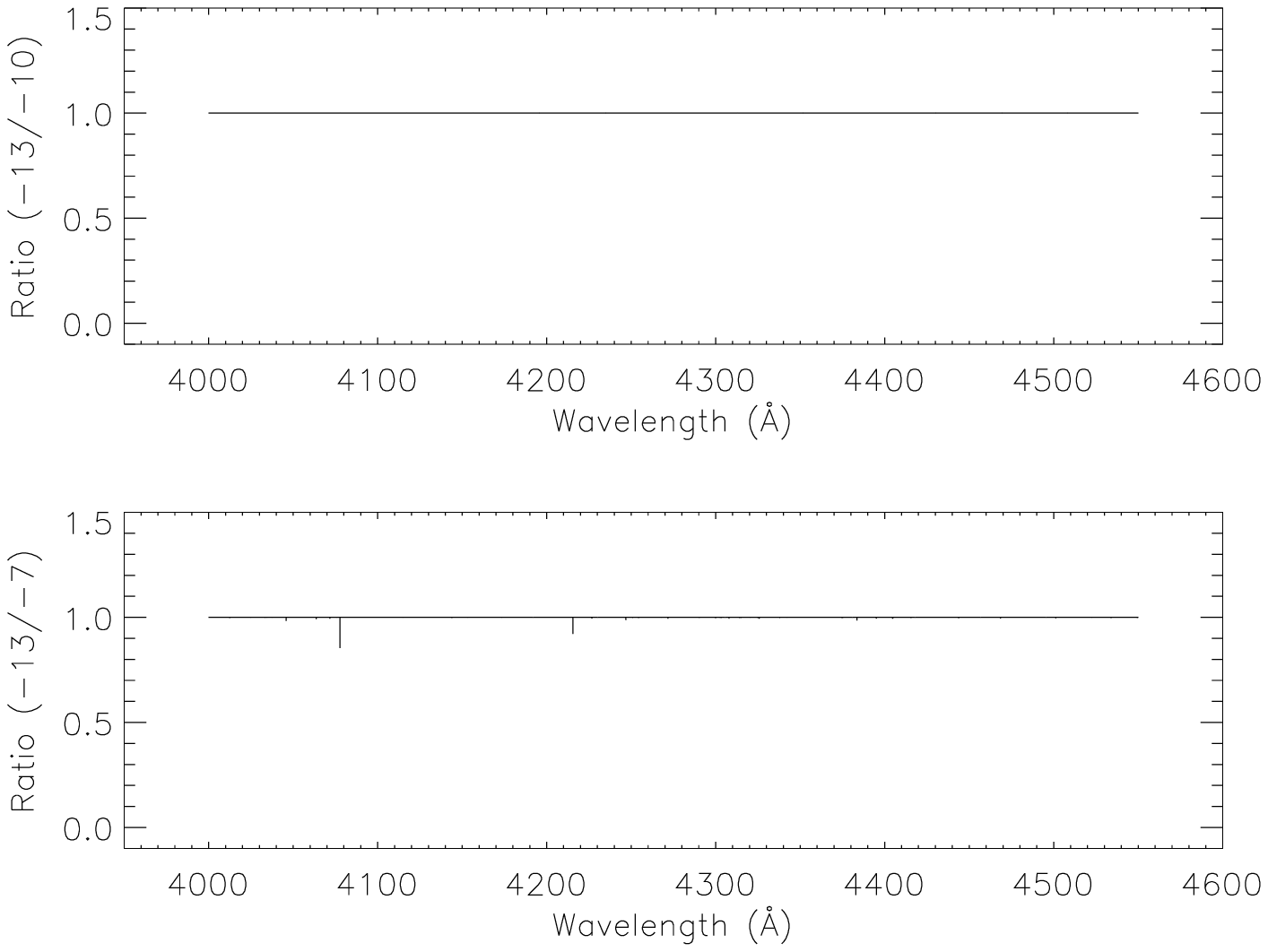}
\caption{Flux ratio of synthetic spectra computed from models with different
$\tau_{\rm Ross}$ limits (see Sec.~\ref{sec:3} for explanation).}
\label{fig:4}
\end{figure}

Another interesting exercise we have conducted is the comparison of the effects on the UV-blue 
spectrum of different parts of the atmosphere. We have segmented the CCS model so as to have 
three models with different limiting optical depth at the surface: $\log{\tau_{\rm Ross}}=-13$, 
$\log{\tau_{\rm Ross}}=-10$, and $\log{\tau_{\rm Ross}}=-7$. For each model we have calculated 
a synthetic spectrum (at $R=522\,000$). Figure~\ref{fig:4} shows the flux ratios in two spectral 
windows, 2000--2550~\AA\ ({\it left panels}) and 4000--4550~\AA\ ({\it right panels}), using the 
flux of the model with limiting depth $\log{\tau_{\rm Ross}}=-13$ as a reference. In the lower 
left panel, where we plot the ratio $F_{\lambda}(\log \tau_{Ross}= -13)/F_{\lambda}
(\log \tau_{\rm Ross}=-7)$, we can visualize the strong effects of the external layers, which 
produce deeper lines. These effects might turn out to be important also in models which include 
non-thermal heating.

\section{Concluding Remarks}
\label{sec:4}
The main result of this work is that {\sc Synthe} series of codes is capable of treating 
the CCS model atmospheres, which reach very low values of $\tau_{\rm Ross}$. The analysis of 
the effects of extending the atmosphere indicates that at mid-UV wavelengths the effects are 
significant while negligible in the blue. The following steps are to extend the analysis to 
models with atmospheric parameters different of the Sun, to complement the opacity (both 
continuous and of lines) as well as to introduce more chemical species. We are in the process of 
including convection for intermediate and cool star models. A detailed comparison at high 
resolution with an observed solar atlas (Kurucz et al. 1984) is also underway.



\end{document}